\documentclass{article}

\usepackage[preprint]{neurips_2025}
\usepackage{algorithm}
\usepackage{algorithmic}

\usepackage[utf8]{inputenc} % allow utf-8 input
\usepackage[T1]{fontenc}    % use 8-bit T1 fonts
\usepackage{hyperref}       % hyperlinks
\usepackage{url}            % simple URL typesetting
\usepackage{booktabs}       % professional-quality tables
\usepackage{amsfonts}       % blackboard math symbols
\usepackage{nicefrac}       % compact symbols for 1/2, etc.
\usepackage{microtype}      % microtypography
\usepackage{xcolor}         % colors

\usepackage{graphicx} % Required for inserting images

\usepackage{amsfonts}
\usepackage{bm}
\usepackage{dsfont}
\usepackage{amsmath}
\usepackage{amsthm}
\usepackage{nicefrac}
\usepackage{bbm}

\usepackage{graphicx}
\usepackage{hyperref}
\usepackage{amssymb}
\usepackage{wrapfig}
\usepackage{caption}
\usepackage{geometry}
\usepackage{subfigure}
\usepackage{subcaption}
\usepackage{cleveref}
\usepackage{xspace}

\usepackage{color}   %May be necessary if you want to color links
\usepackage{graphicx}
\usepackage{hyperref}
\hypersetup{colorlinks=true,linkcolor=blue}

\newcommand{\actiongraph}{\ensuremath{\mathsf{ActionGraph}}\xspace}

\title{Representation of Inorganic Synthesis Reactions and Prediction: Graphical Framework and Datasets}

\author{
  Samuel Andrello\\
  Columbia University\\
  New York, NY 10027\\
  \And
  Daniel Alabi\\
  University of Illinois at Urbana-Champaign\\
  Urbana, IL 61801\\
  \And
  Simon J. L. Billinge\\
  Columbia University\\
  New York, NY 10027\\
}

\begin{document}

\maketitle

\begin{abstract}
While machine learning has enabled the rapid prediction of inorganic materials with novel properties, the challenge of determining how to synthesize these materials remains largely unsolved. Previous work has largely focused on predicting precursors or reaction conditions, but only rarely on full synthesis pathways. We introduce the \actiongraph, a directed acyclic graph framework that encodes both the chemical and procedural structure, in terms of synthesis operations, of inorganic synthesis reactions. Using 13,017 text-mined solid-state synthesis reactions from the Materials Project, we show that incorporating PCA-reduced \actiongraph adjacency matrices into a $k$-nearest neighbors retrieval model significantly improves synthesis pathway prediction. While the \actiongraph framework only results in a 1.34\% and 2.76\% increase in precursor and operation F1 scores (average over varying numbers of PCA components) respectively, the operation length matching accuracy rises 3.4 times (from 15.8\% to 53.3\%). We observe an interesting trade-off where precursor prediction performance peaks at 10-11 PCA components while operation prediction continues improving up to 30 components. This suggests composition information dominates precursor selection while structural information is critical for operation sequencing. Overall, the \actiongraph framework demonstrates strong potential, and with further adoption, its full range of benefits can be effectively realized.

\end{abstract}

\section{Introduction}\label{sec:intro}

Inorganic solid-state synthesis is notoriously challenging to predict and design due to the lack of a general theory describing how phases form and transform under given conditions. In particular, there is no comprehensive theoretical framework dictating how precursor phases evolve into products during heating and other processing steps \cite{elsamman2024}. As a result, synthesis routes are often developed by trial-and-error or past experience. Moreover, real-world data on multi-step syntheses with intermediate phases is scarce\cite{alabi2024empiredbdataacceleratecomputational}. Most reported synthesis recipes focus only on the starting materials and final products, omitting the transient \textit{intermediate} phases that may appear along the reaction pathway. Even with recent efforts to compile text-mined datasets of inorganic synthesis recipes \cite{kononova2019}, the available data remain limited in capturing the full complexity of synthesis processes (especially intermediate steps). This combination of a knowledge gap and limited data motivates new computational approaches to guide inorganic materials synthesis.

In this work, we address the problem of identifying which \textit{precursor} materials and intermediate steps can lead to the desired \textit{products} in inorganic synthesis, given only the target products as input. We propose a novel graph-based framework, termed the \textit{\actiongraph}, to represent an entire inorganic synthesis pathway. Using this representation, we employ $k$-nearest neighbors models~\citep{Cover13} with dimensionality-reduced graph adjacency matrices to predict the reaction steps-specifically, to predict likely precursors and intermediates that yield the target product Our ultimate goal is to develop a predictive model that can suggest how to synthesize new materials by learning from known examples. Such a model could significantly reduce the trial-and-error in materials discovery and lay the groundwork for automated synthesis planning. Indeed, data-driven methods are beginning to play a role in rationalizing and predicting synthesis outcomes \cite{huo2022}, and an accurate precursor-prediction model could eventually be integrated into robotic synthesis laboratories for closed-loop experimentation.

\begin{figure}[!htbp]
    \centering
    \begin{minipage}{0.48\textwidth}
    \centering
    \includegraphics[width=\linewidth]{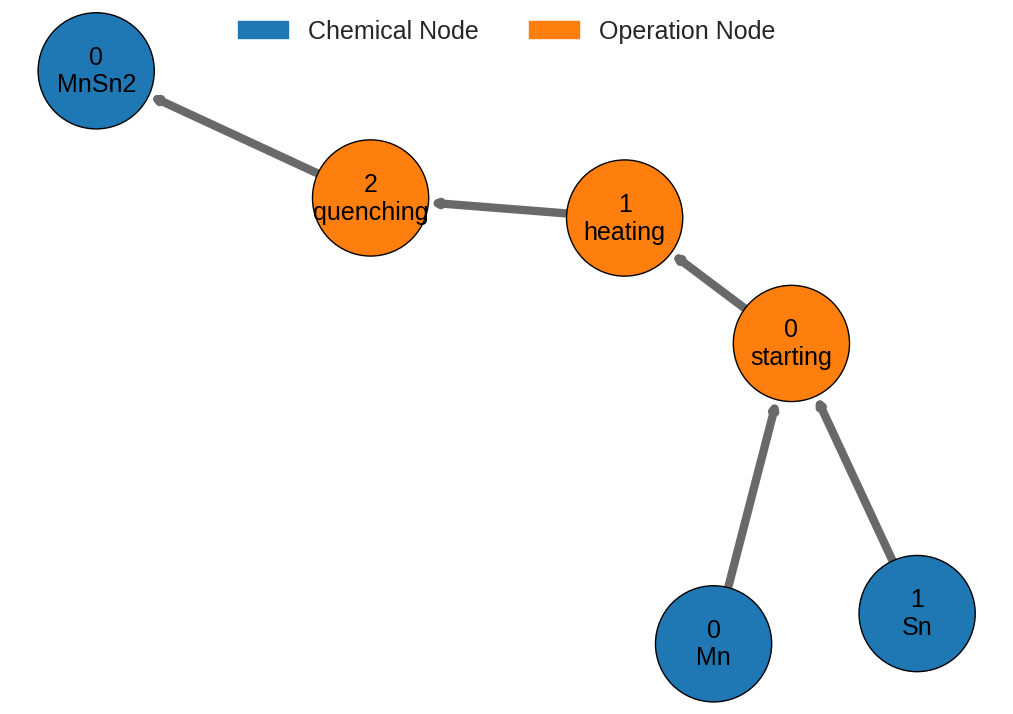}
    \caption*{(a) Mn + 2Sn $\rightarrow$ MnSn$_2$}
    \end{minipage}
    \hfill
    \begin{minipage}{0.48\textwidth}
    \centering
    \includegraphics[width=\linewidth]{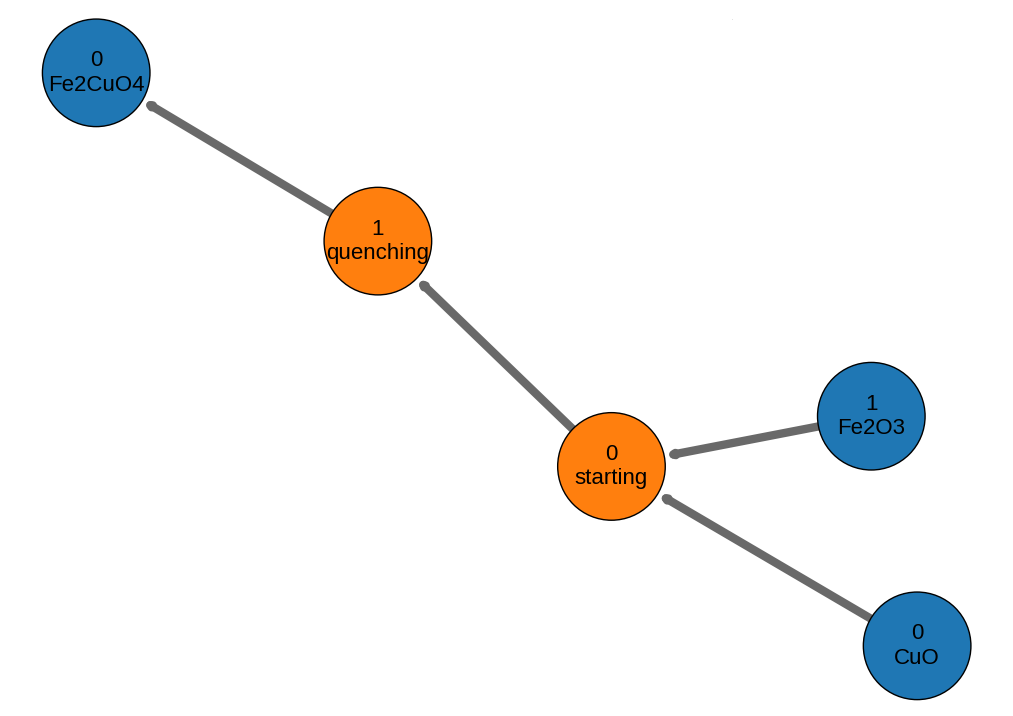}
    \caption*{(b) CuO + Fe$_2$O$_3$ $\rightarrow$ Fe$_2$CuO$_4$}
    \end{minipage}
    \caption{Two synthesis reactions represented in \actiongraph form. Nodes are labeled with their corresponding indices. Chemical nodes placed prior to operations have a separate set of indices than chemical nodes placed after operations.}
    \label{fig:synthesis-ex}
\end{figure}

\subsection{Inorganic Synthesis: Why is the problem difficult?}

Inorganic synthesis presents unique challenges compared to organic synthesis, making it particularly difficult to predict and automate. While organic synthesis benefits from established retrosynthesis frameworks~\cite{wei_machine_2025} that allow for systematic planning of reaction pathways, inorganic synthesis lacks analogous theoretical foundations. This absence of a comprehensive theory means that predicting how precursors transform into products during solid-state reactions remains largely empirical and experience-driven.

Several factors contribute to this complexity. First, inorganic syntheses often involve high-temperature solid-state reactions where atomic diffusion and phase transformations occur through mechanisms that are difficult to observe directly. Second, reaction pathways frequently involve transient intermediate phases that may not be reported in the literature or captured in databases. Third, the outcome of an inorganic synthesis depends heavily on processing conditions such as temperature profiles, atmospheres, and mechanical treatments, creating a vast parameter space that is challenging to navigate systematically.

This complexity has significant implications for materials innovation. For instance, the development of novel battery materials~\cite{szymanski_toward_2021} often involves extensive trial-and-error experimentation to discover viable synthesis routes, substantially slowing the transition from computational material design to physical realization. The field has consequently seen growing interest in "closed-loop" synthesis approaches~\cite{szymanski_toward_2021}, where automated experimentation platforms could iteratively refine synthesis protocols with minimal human intervention.

Unlike organic chemistry, where retrosynthesis provides a structured method to solve the synthesis inverse problem (finding precursors and conditions that yield a target molecule), inorganic materials science lacks such systematic frameworks. This gap has motivated recent efforts to develop data-driven approaches that can learn from existing synthesis recipes to predict viable pathways for novel materials~\cite{kim_predicting_2024, mcdermott_graph-based_2021}. The \actiongraph framework we present aims to address this challenge by providing a structured representation of inorganic synthesis that captures both the chemical constituents and the procedural aspects of materials preparation.

Our contributions are as follows:
\begin{itemize}
\item The \actiongraph framework which in this paper is being used to represent inorganic synthesis reactions, but could be extended to other processes or any set of actions with a process-like structure.
\item A dataset of 13,017 \actiongraph graphs transformed from open-source Materials Project data, each representing a single solid-state inorganic synthesis reaction.
\end{itemize}

Next, we discuss some related work.

\section{Related Work}

\subsection{Machine Learning for Sciences}

Our work on \actiongraph sits in the domain of research to further enable
scientific discovery via the use of machine learning.
Recently, the application of machine learning (ML) to scientific domains has grown rapidly. Such applications have led to new and more efficient
simulations and has facilitated the discovery of patterns in complex, high-dimensional data. For example, in the physics community, ML has been employed to uncover hidden structures in quantum systems~\citep{carleo2017solving, mills2021role}, model phase transitions~\citep{nieuwenburg2017learning}, and approximate solutions to differential equations~\citep{raissi2019physics}. In chemistry and materials science, deep learning techniques have achieved impressive performance in molecular property prediction~\citep{gilmer2017neural}, drug discovery~\citep{zhavoronkov2019deep}, and the generation of novel molecules~\citep{gomez2018automatic}. However, none of these works develop
frameworks nor datasets to address the important task of synthesis of
materials. Our work aims to bridge this gap.

In terms of ML architectures,
Graph neural networks (GNNs) or graph-based architectures have emerged as a key tool for modeling structured scientific data. They have been particularly effective in learning interatomic potentials~\citep{schutt2018schnet}, modeling physical interactions~\citep{battaglia2018relational, BP24}, and predicting protein structures~\citep{jumper2021highly}. Meanwhile, generative models, including variational autoencoders and generative adversarial networks, have facilitated the inverse design of materials~\citep{sanchez2018inverse} and molecular synthesis~\citep{popova2018deep}.
Recent work also explores how ML can augment simulation, such as learning surrogate models for climate prediction~\citep{rolnick2019tackling} or accelerating Monte Carlo methods in statistical physics~\citep{liu2017accelerated}. Furthermore, 
Reinforcement learning has been applied to optimize experimental design and automate laboratory processes~\citep{macleod2020self}.

Despite these advances, challenges remain in developing ML methods that incorporate physical constraints, provide uncertainty estimates, and generalize beyond observed regimes~\citep{BL07, 2008AcCrA..64..631J, NZTB25}. Our work on \actiongraph presents a novel
graph-based framework to capture the problem of inorganic synthesis and to
incorporate physical constraints. We
also provide datasets to evaluate the efficacy of predicting synthesis
pathways.

\subsection{Synthesis Prediction}

Synthesis prediction has evolved along different trajectories in organic and inorganic chemistry, reflecting the distinct challenges in each domain. In organic chemistry, retrosynthetic analysis-working backward from target molecules to starting materials through known reaction types-has been formalized for decades and recently enhanced through machine learning approaches~\cite{wei_machine_2025}. These methods benefit from large, structured datasets of reactions and well-established reaction mechanisms.

In contrast, inorganic synthesis prediction has emerged more recently and faces unique challenges. McDermott et al.~\cite{mcdermott_graph-based_2021} introduced a graph-based framework to predict solid-state reaction pathways, focusing on thermodynamic driving forces rather than synthesis procedures. Their approach, while pioneering, primarily addresses reaction energetics rather than practical synthesis protocols.

He et al.~\cite{he_precursor_2023} developed a machine learning approach specifically for precursor recommendation, achieving significant accuracy in predicting starting materials for inorganic syntheses. However, their work focuses primarily on precursor selection rather than complete synthesis pathways including operations and conditions.

More recently, Kim et al.~\cite{kim_predicting_2024} proposed an elementwise template formulation for predicting inorganic synthesis recipes, while another study by Kim et al.~\cite{kim_large_2024} explored using large language models for inorganic synthesis predictions. Recently, Noh et al.~\cite{noh2025} predicted reaction components using an attention mechanism. These approaches make important strides but focus primarily on either precursors or reaction conditions rather than the complete synthesis workflow.

Karpovich et al.~\cite{karpovich_inorganic_2021} addressed reaction condition prediction using generative machine learning, emphasizing the parameter space of synthesis conditions rather than the structural aspects of synthesis pathways.

What distinguishes our work is the \actiongraph framework's holistic representation of inorganic synthesis as a directed process flow capturing both materials and operations. Unlike previous approaches that focus on either precursors or conditions in isolation, our methodology encodes the entire synthesis pathway as a graph structure, enabling prediction of complete synthesis routes rather than individual components.

Now, we provide preliminary definitions for our results.

\section{Preliminaries}\label{sec:preliminaries}
% \sjb{I started the work of ``skeletonizing" the text to make it easier to move text around.  In general, we prefer skels that are short descriptions of what needs to be said in paragraph in just one or two lines, but told as if it is written as part of the story. @Sam, we may want to try and start over writing the skeleton and copy-paste some of these bigger blocks of text back as we need them?  That would be my suggestion.}

\textbf{\actiongraph definition.} We model an inorganic synthesis process as an \emph{\actiongraph}, which is a directed, acyclic graph capturing the relationships between starting materials, intermediate compounds, and final products through the sequence of synthesis actions. Formally, an \actiongraph $G=(V,E)$ consists of nodes $V$ representing chemical compounds (phases) and operations (synthesis steps). The directed edges $E$ represent "action flow," or the flow of phases through the synthesis process. The graph is acyclic, reflecting the temporal order of synthesis steps.

\textbf{Precursors and products.} In an \actiongraph, we distinguish two roles for compound nodes. \textit{Precursors} are the initial reactant materials provided at the start of the synthesis (these correspond to source nodes with no incoming edges in the graph). \textit{Products} are the final target materials of the synthesis, which only have incoming edges. For example, if a recipe starts with powders of $A$ and $B$ (precursors), mixes and heats them to form an intermediate compound $C$, and upon further heating yields final product $D$, we would represent this as precursor nodes $A, B$ feeding into intermediate node $C$ (via a ``heating'' edge), which in turn connects to product node $D$ via another action edge.

\textbf{Operations.} In order to capture the nature of synthesis, an \actiongraph also contains operation nodes which represent the actual synthesis steps that transform precursors into products. They are topologically ordered after precursor nodes, but before product nodes. These nodes can represent operations such as heating, mixing, and milling that reflect a high-level view of a syntehsis process (i.e. what is actually being done to the precursors and reaction intermediates).

\textit{Formal definition.}
Define an \actiongraph as $G = (V, E)$ with

$V = C \cup O$ where $C$ is the set of all chemical nodes and $O$ is the set of all operation nodes,

$E \subset V \times V$ is the set of directed edges representing the flow of the synthesis process.

Define a chemical node $C_i = \{\mathrm{element}_i : \mathrm{subscript}_i \}^{m}_{i=1}$ where $\mathrm{subscript}_i$ is the stoichiometric coefficient for $\mathrm{element}_i$.

Define $C_{in}$ and $C_{out}$, the sets of all precursors and products, respectively. Then $C \equiv C_{in} \cup C_{out}$.

Define an operation node as

$O_i = \left( \mathrm{type}_i, \ \{\mathrm{condition}_j : \mathrm{value}_j \}^{m}_{j=1}, \ \mathrm{metadata}_j \right)$

where $\mathrm{type}_i \in \{ \mathrm{StartingSynthesis}, \mathrm{MixingOperation}, \mathrm{ShapingOperation}, \\\\\mathrm{DryingOperation}, \mathrm{HeatingOperation}, \mathrm{QuenchingOperation} \}$.

\textit{Structural constraints.} We also consider some constraints on the structure of an \actiongraph that are required for its validity:
\begin{enumerate}
    \item $\forall c \in C_{in}, \ \mathrm{indegree}(c) = 0 \iff$ precursor nodes have no incoming edges.
    \item $\forall c \in C_{out}, \ \mathrm{outdegree}(c) = 0 \iff$ product nodes have no outgoing edges.
    \item $\forall o \in O, \ \mathrm{indegree}(o) \geq 1$ and $\mathrm{outdegree}(o) \geq 1 \iff$ operation nodes have at least one input and one output.
    \item $G$ is acyclic.
\end{enumerate}

\textbf{Synthesis workflow and modeling challenges.} A typical solid-state synthesis workflow involves several sequential actions: precursors are first combined (mixed, ground together, sometimes pressed or shaped into pellets), then subjected to thermal treatments such as drying (to remove solvents or binders) and one or more heating steps (calcination, sintering) at high temperatures, and possibly cooling or quenching steps at the end. These processing actions can lead to the formation of new crystalline phases as intermediates before the final product phase is obtained. Modeling this process computationally is challenging because the outcome of each step can depend non-linearly on the combination of precursors, the specific conditions (temperature, time), and the sequence of actions. The \actiongraph framework provides a structured way to encode all these aspects (materials and actions) into a single representation. However, learning to predict the sequence from precursors to product requires capturing potentially complex relationships in the graph (e.g., how multiple precursor compounds might combine to form a particular intermediate). The lack of abundant, richly annotated data detailing intermediate information such as crystal structure and reaction conditions further complicates direct modeling. In addition, it is desirable to capture the high-level synthesis process from this data, as the goal is to predict synthesis pathways that are directly usable in a laboratory setting. These challenges necessitate an adaptable representation for synthesis (hence our graph-based approach) along with benchmarks on the available data.

\section{Methodology}\label{sec:methodology}

%\SA{REQUIRED: Add reference to publicly available code and data}

\subsection{Initial Dataset Construction}

\textbf{Dataset and preprocessing.} We leverage a recently published dataset of inorganic materials synthesis recipes text-mined from the literature \cite{kononova2019}. This dataset (from the Materials Project repository) contains thousands of solid-state reaction examples, each including a list of precursors, details of synthesis operations (e.g. “heat at 800$^\circ$C for 10 hours”), and the resulting product (and occasionally intermediate phases if reported by the authors). The initial, unfiltered set numbered 30,850 JSON files, each representing a single solid-state synthesis reaction. To focus on well-defined reactions, we filtered out any entries with variable or non-stoichiometric compositions (e.g. recipes that involve formula variables or unspecified proportions), as these are difficult to interpret in a standardized way. Furthermore, erroneous reactions, or those that contain invalid or empty chemical formula, were also removed. The final dataset numbered 13,017 reactions.

After cleaning, a second dataset was constructed by transforming each synthesis reaction into our \actiongraph representation. In this conversion, the precursors are set as the initial nodes, any operations as nodes following this, and the products as the final nodes. Edges are added to connect these nodes in the order of the described steps. For instance, if a paper describes that mixing compounds $\text{A}$ and $\text{B}$ followed by heating yields intermediate $\text{C}$, which upon further heating yields product $\text{D}$, we create edges $A \rightarrow C$, $B \rightarrow C$ (for the mixing+heating step forming $C$) and $C \rightarrow D$ (for the subsequent step yielding $D$). Each edge is annotated with the type of action (mixing, heating, etc.), though in our current model we focus primarily on the graph connectivity and not the exact conditions due to the lack of sufficient data. This second dataset was of the same length as the untransformed, filtered dataset: 13,017 serialized \actiongraph objects.

Data sharing in materials science is not encouraged due to
copyright or legal reasons~\citep{AGMSW25}. As a result, we can
rely on publicly-available datasets instead.

\textbf{Approach overview.} 

In order to demonstrate the utility of the \actiongraph framework, we evaluate its performance through comparative experiments on two distinct datasets. The first dataset is derived from the filtered Materials Project (MP) data, which provides material compositions and associated properties in a conventional tabular format. Using this dataset, we construct a baseline $k$-nearest-neighbors ($k$-NN) model with $k = 1$. This model serves as a simple, interpretable benchmark that relies purely on standard feature vectors extracted directly from the material property data, without any graph-based augmentation.

Next, to assess the added value introduced by the \actiongraph framework, we train a second $k$-NN model using data processed through \actiongraph. This second dataset encodes domain knowledge and inter-entity dependencies by representing materials and reactions as nodes and edges in a graph. To incorporate this graph structure into the learning pipeline, we transform each data instance into an enriched feature vector that includes not only conventional material attributes, but also information derived from the graph topology. Specifically, we integrate structural features obtained from the adjacency matrix of the \actiongraph, which encodes the connectivity patterns among nodes (e.g., reactions and precursors). This embedding captures relational and contextual cues, such as shared precursors, co-occurring reactions, or common synthesis pathways, that are otherwise absent in flat representations. By augmenting the feature space in this manner, we enable the second $k$-NN model to exploit both local material properties and higher-order structural correlations, thus providing a more holistic assessment of material similarities and model performance within the \actiongraph paradigm.

\textbf{Featurization.} Each synthesis reaction is featurized to serve as input to a knn. In our preliminary implementation, we use a simple composition-based feature vector: essentially a representation of the chemical formula of the compound. This can be, for example, an element fraction vector indicating the proportion of each element in the compound's formula, or a one-hot encoding of the compound identity if it is recognized from a known list. The intuition is to provide the model with information about the chemical makeup of each node (since knowing the elements present in a precursor vs in the product might help the model learn which precursors could lead to that product). this composition vector is then concatenated with another feature vector containing the averages of each of atomic mass, atomic radius, melting point, Pauling electronegativity, electron affinity, and ionization energy for each chemical. However, this straightforward featurization may not capture more subtle aspects of synthesis and is limited by the lack of structural characterization information of chemicals in the dataset. Exploring more sophisticated featurization (incorporating, e.g., known chemical descriptors or features learned from materials graphs \cite{elsamman2024}) is an area of ongoing work.

\begin{figure}[!htbp]
  \centering
  \begin{minipage}{0.31\textwidth}
    \centering
    \includegraphics[width=\linewidth]{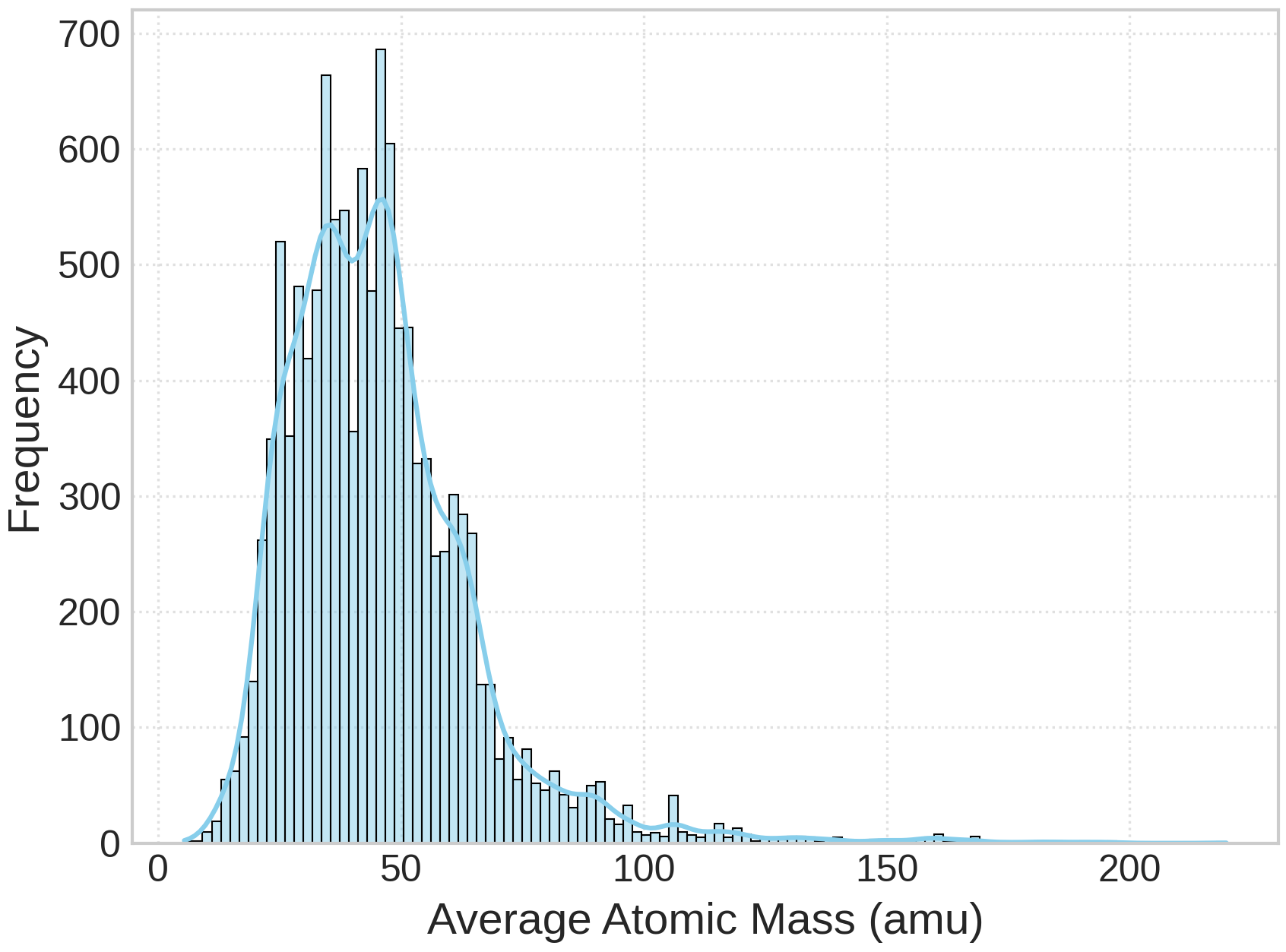}
    \caption*{}
  \end{minipage}\hfill
  \begin{minipage}{0.31\textwidth}
    \centering
    \includegraphics[width=\linewidth]{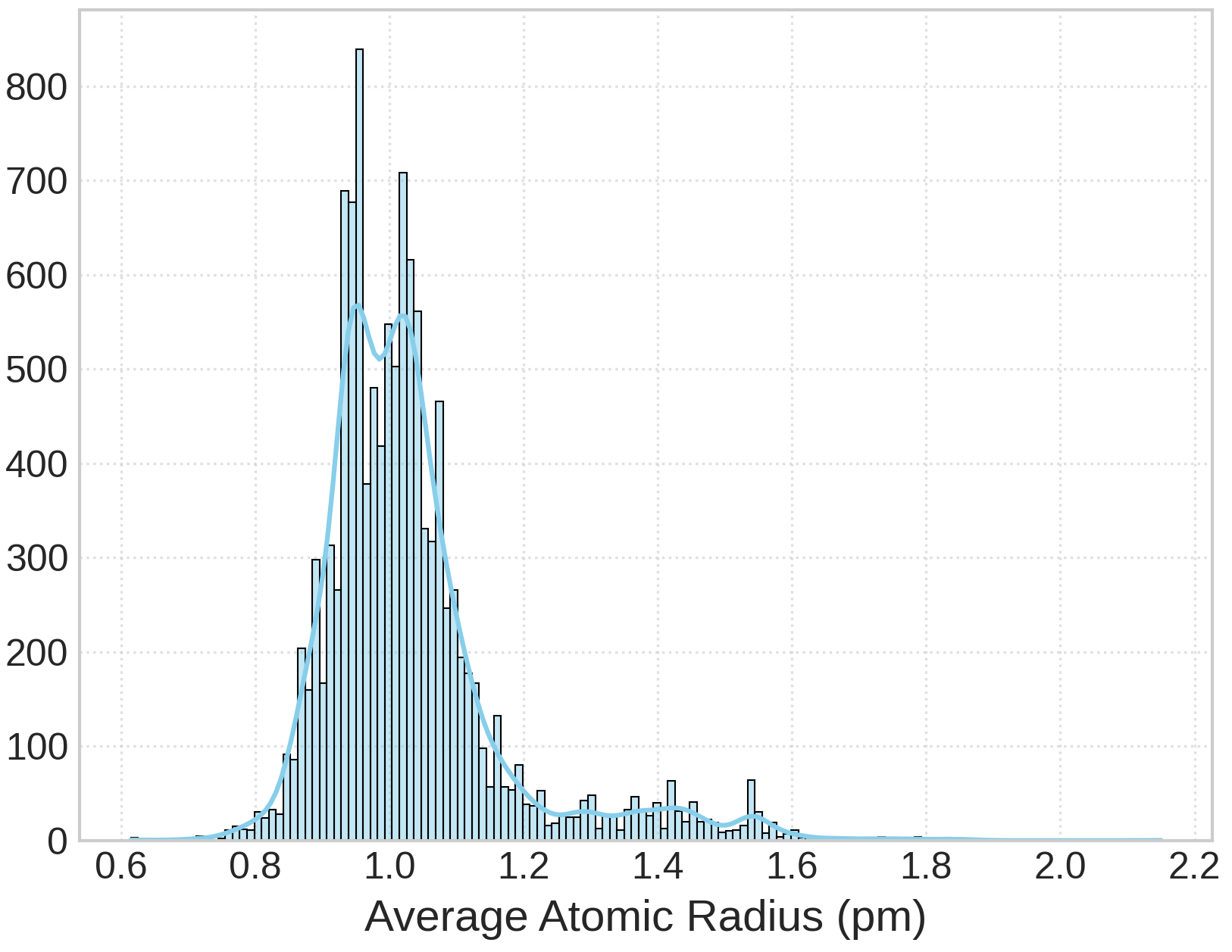}
    \caption*{}
  \end{minipage}\hfill
  \begin{minipage}{0.31\textwidth}
    \centering
    \includegraphics[width=\linewidth]{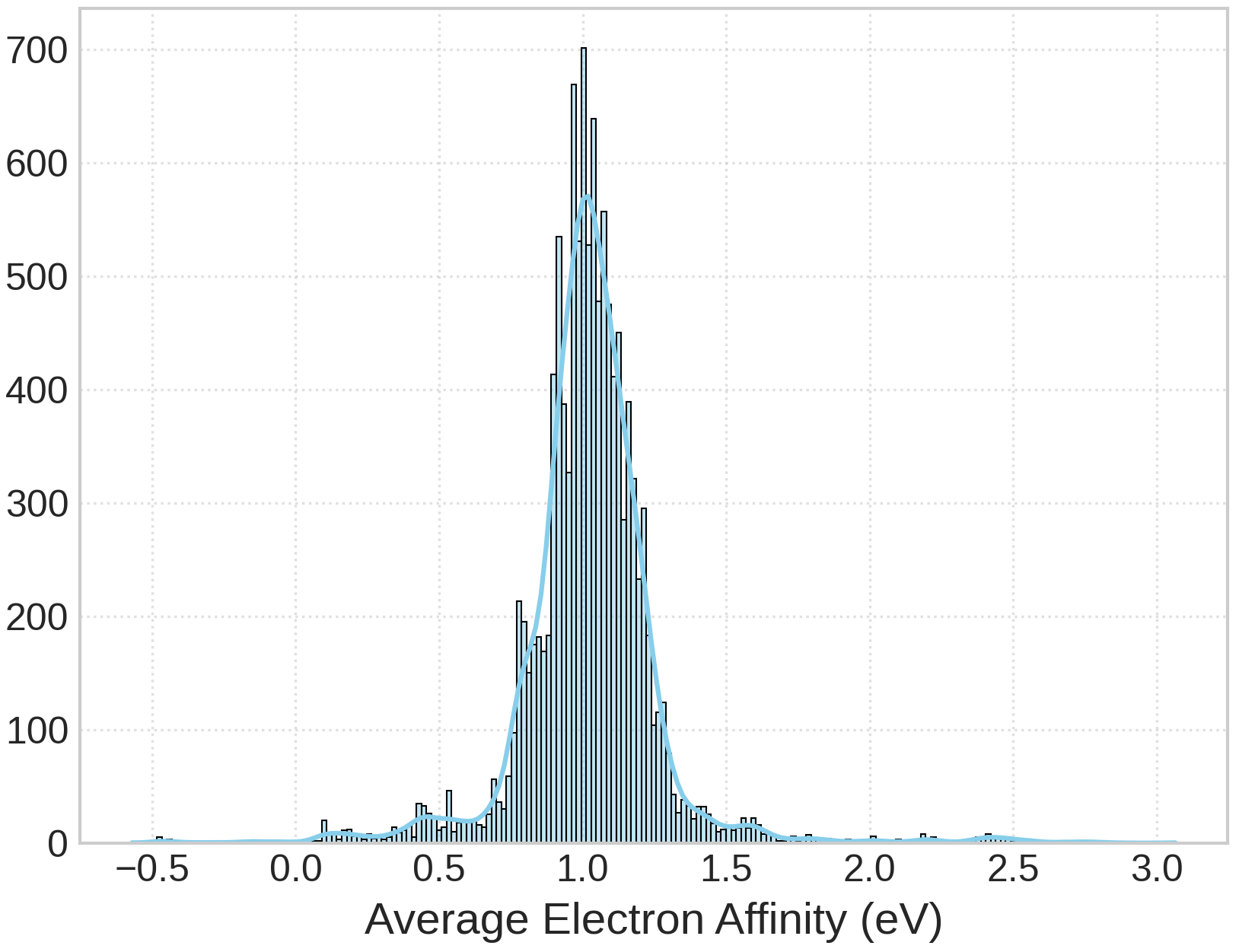}
    \caption*{}
  \end{minipage}
  
  \vspace{-1.2em}
  
  \begin{minipage}{0.31\textwidth}
    \centering
    \includegraphics[width=\linewidth]{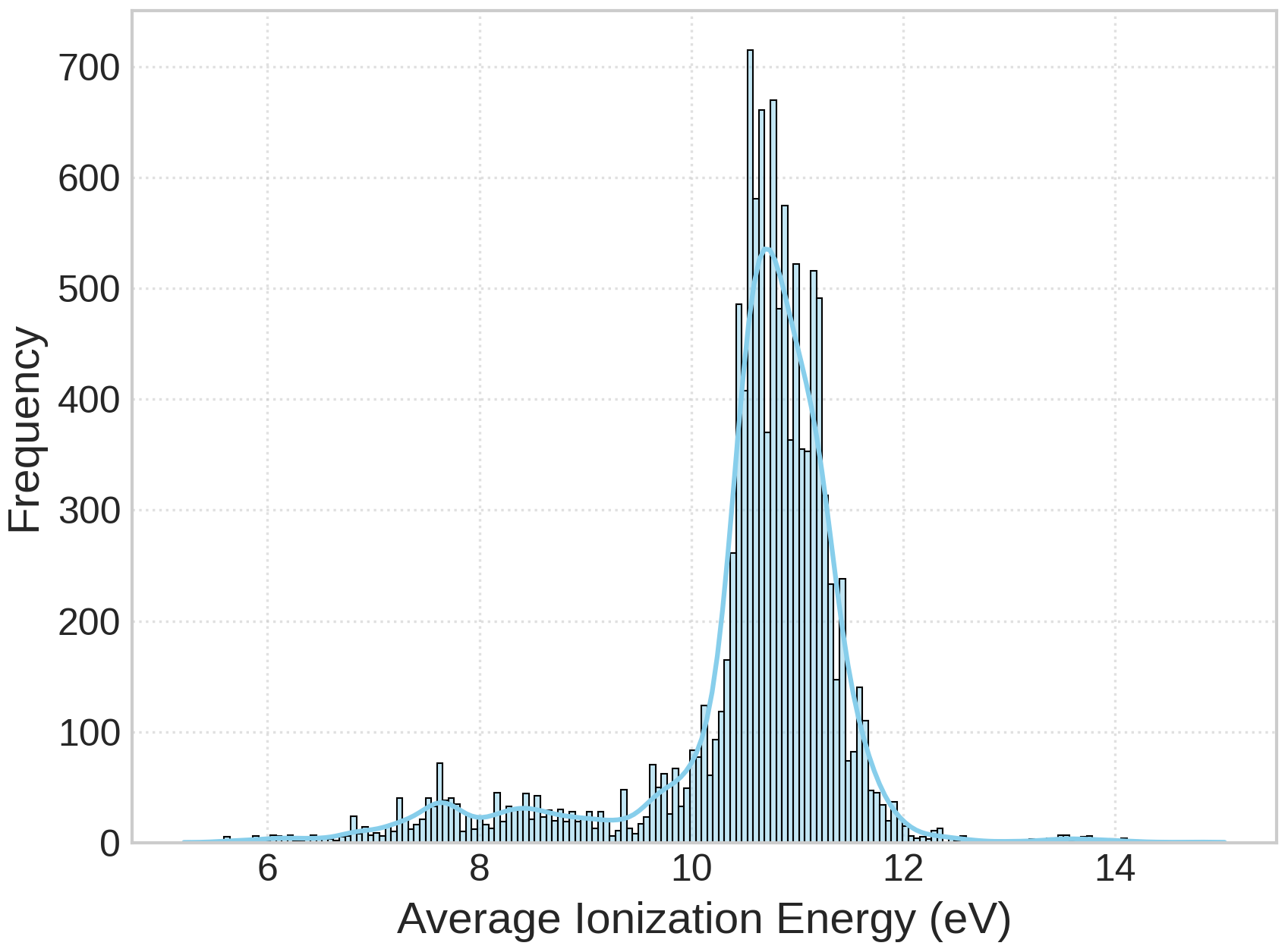}
    \caption*{}
  \end{minipage}\hfill
  \begin{minipage}{0.31\textwidth}
    \centering
    \includegraphics[width=\linewidth]{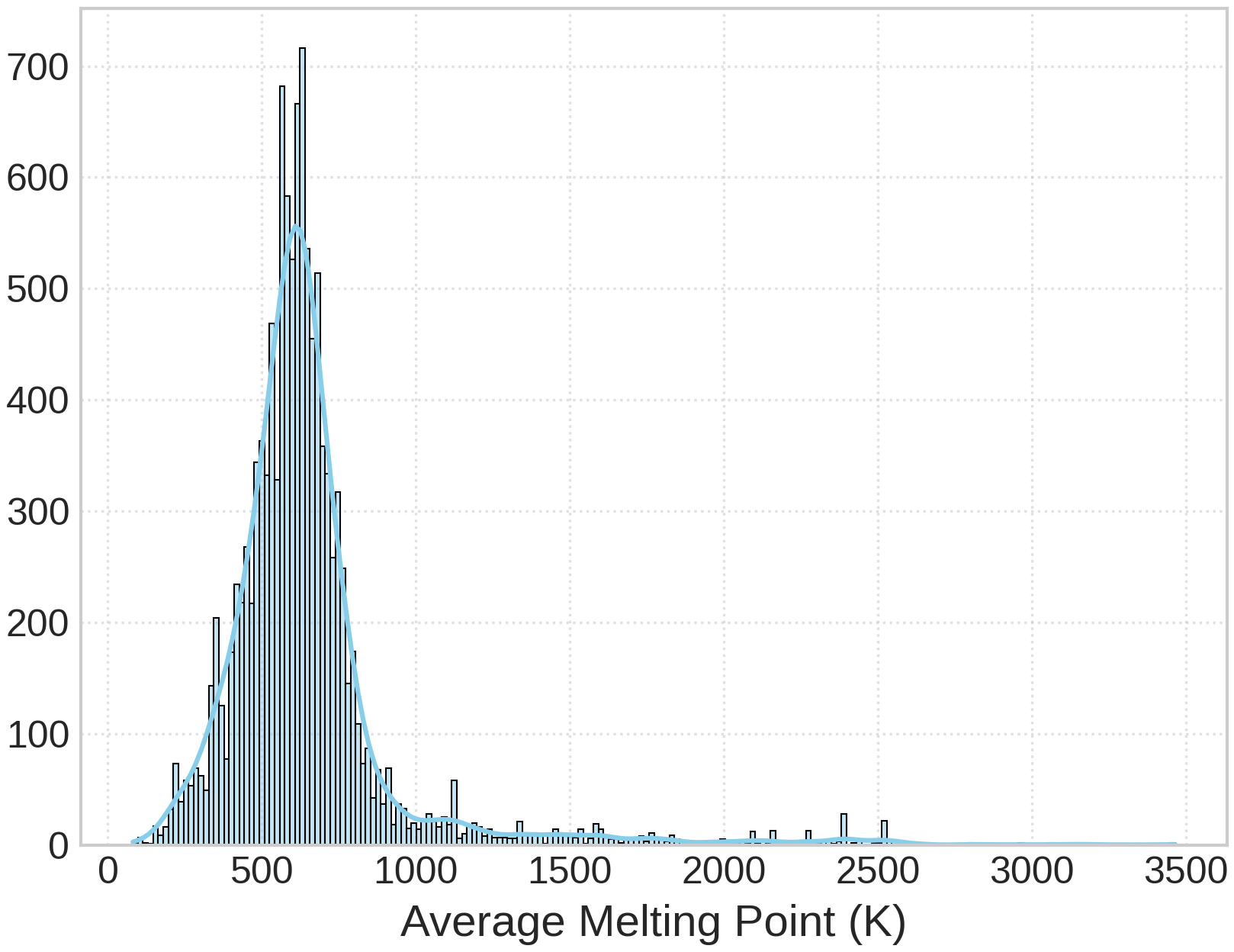}
    \caption*{}
  \end{minipage}\hfill
  \begin{minipage}{0.31\textwidth}
    \centering
    \includegraphics[width=\linewidth]{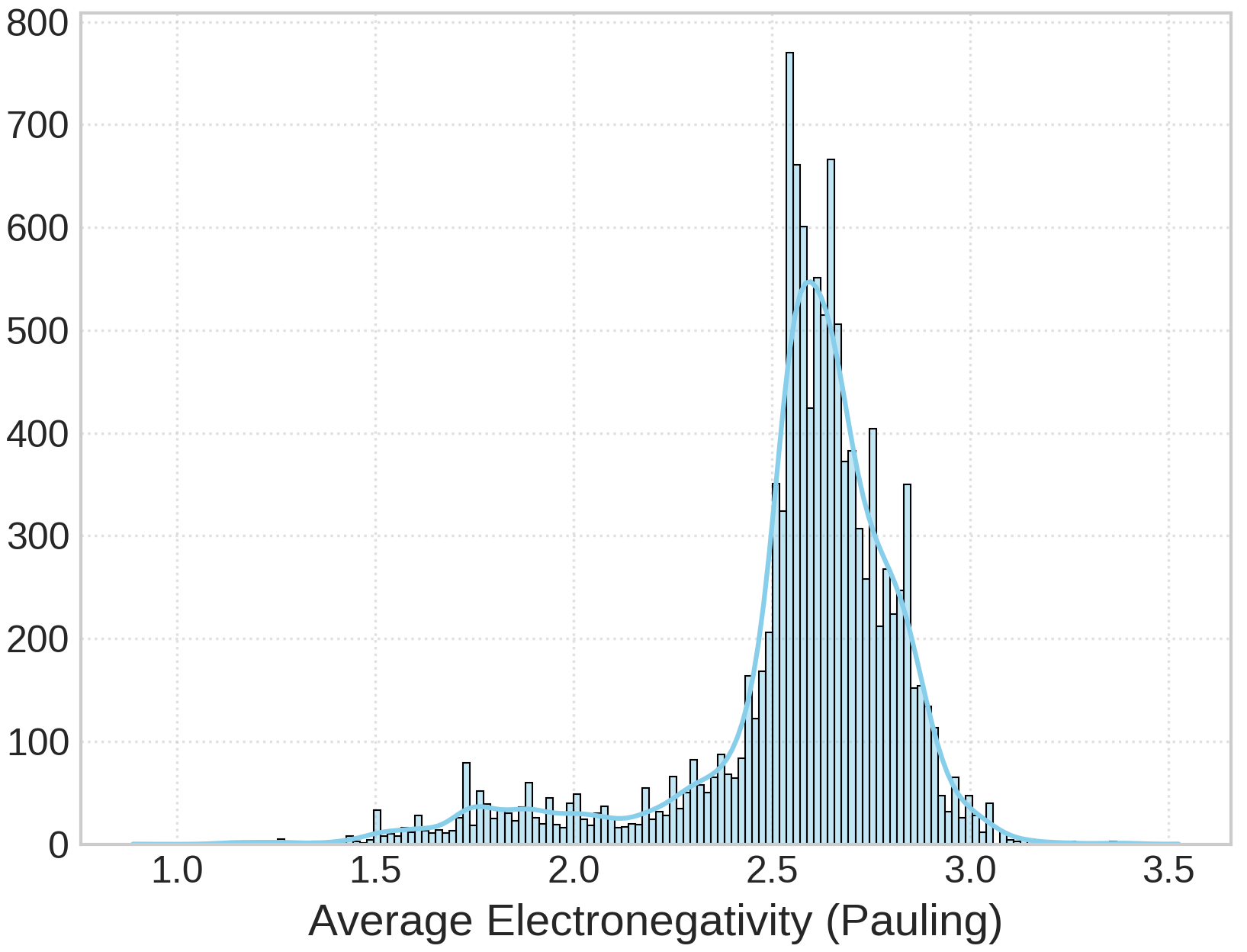}
    \caption*{}
  \end{minipage}
  \vspace{-2em}
  \caption{Property distributions of the featurized dataset. Frequency represents the number of total chemicals that contain a property with a value within that bin. Each property value is the average of the elemental properties for the composition of a specific chemical.}
\end{figure}

\textbf{Utilizing the \actiongraph.} In order to assess the impact of the topology of the synthesis reaction on predicting precursors and operations, we instead train a $k$-NN model on the second dataset. The \actiongraph structure is incorporated in featurization via a multi-step process:
\begin{enumerate}
    \item Find the maximum number of nodes present in all of the \actiongraph graphs. This was determined to be 31.
    \item During featurization of a target \actiongraph, the adjacency matrix is extracted then padded to a 31-by-31 matrix, flattened, then scaled using \textsc{StandardScaler}.
    \item Principal component analysis (PCA) is applied to reduce the dimensionality of the adjacency matrix such that the ``adjacency matrix vector'' for each \actiongraph is the same length. This is then concatenated with the existing feature vector, which was determined the same way as before.
\end{enumerate}

While this method could suffer from being too simple given the complexity of inorganic synthesis, it provides a straightforward approach for examining the effectiveness of using the \actiongraph without requiring complex neural network architectures.

\textbf{Model architecture.} For both datasets, a $k$-NN model  with $k = 1$ is used, acting as a nearest-analog retrieval system. Specifically, the \textsc{NearestNeighbors} model from \textsc{scikit-learn} is employed with cosine distance as a similarity metric \cite{scikit-learn}. This approach treats synthesis prediction as an information retrieval problem: when queried with a target material, the model identifies the most similar synthesis reaction in the training set set and adopts its complete recipe as the prediction for the new material.

The feature vectors for both models undergo scaling using \textsc{scikit-learn}'s \textsc{QuantileTransformer} with  output\_distribution=``normal" and n\_quantiles=min(1000, \textsc{X\_train}). This nonlinear transformation maps the original feature distributions to follow a normal distribution, which helps improve model performance by reducing the impact of outliers and ensuring consistent feature scales. This is notably important for the elemental properties as different units cause significant differences in their feature scales.

A 75-25 train-test split was used for all experiments to ensure consistent evaluation across the different model configurations. All experiments were performed in a Python $3.12.9$ environment with scikit-learn $1.61$.

\section{Experimental Validation}\label{sec:results}

\textbf{Evaluation metrics.} We adopt several metrics from synthesis prediction literature to evaluate our models. F1 score measures the harmonic mean of precision and recall for precursor and operation prediction, providing a balanced assessment of prediction accuracy. Exact Match (EM)~\cite{kim_large_2024} is a strict metric that counts a prediction as correct only when the entire set of predicted precursors exactly matches the reference set, with no missing or extra compounds. Operation Length Matching~\cite{karpovich_inorganic_2021} measures the percentage of predictions where the number of synthesis operations matches the ground truth, reflecting the model's ability to capture the procedural complexity of synthesis.

To ensure robust evaluation, we employed a 75-25 train-test split of our dataset, resulting in 9,763 samples for training and 3,254 for testing. All reported metrics are computed on the test set. Feature scaling was performed using scikit-learn's QuantileTransformer to ensure robustness to outliers and non-Gaussian distributions in the chemical property features.

\subsection{Principal Component Analysis and Feature Space Visualization}

\begin{figure}[!htbp]
    \centering
    \begin{minipage}{0.46\textwidth}
    \centering
    \includegraphics[width=\linewidth]{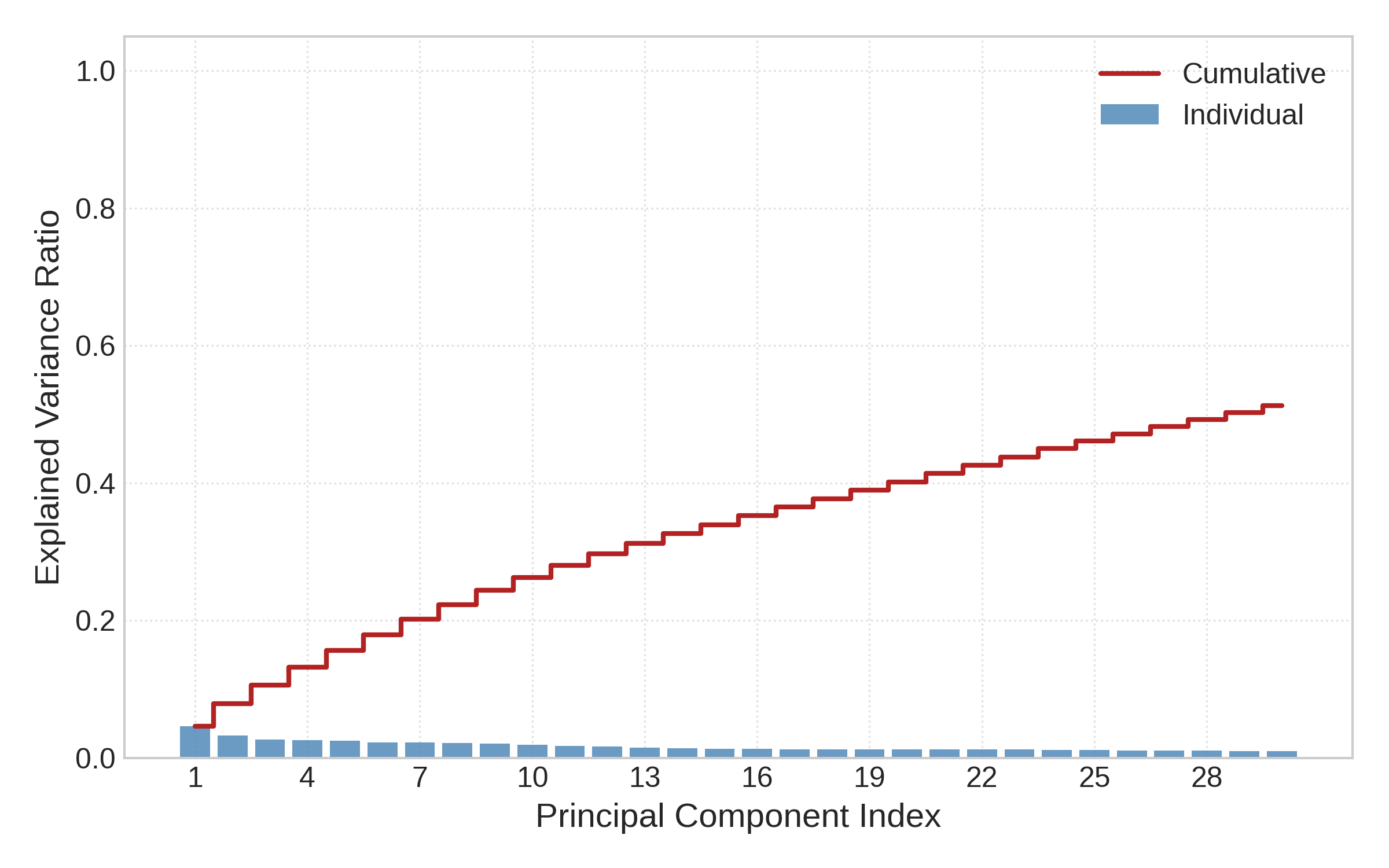}
    \caption*{(a) PCA explained variance ratio}
    \end{minipage}
    \hfill
    \begin{minipage}{0.46\textwidth}
    \centering
    \includegraphics[width=\linewidth]{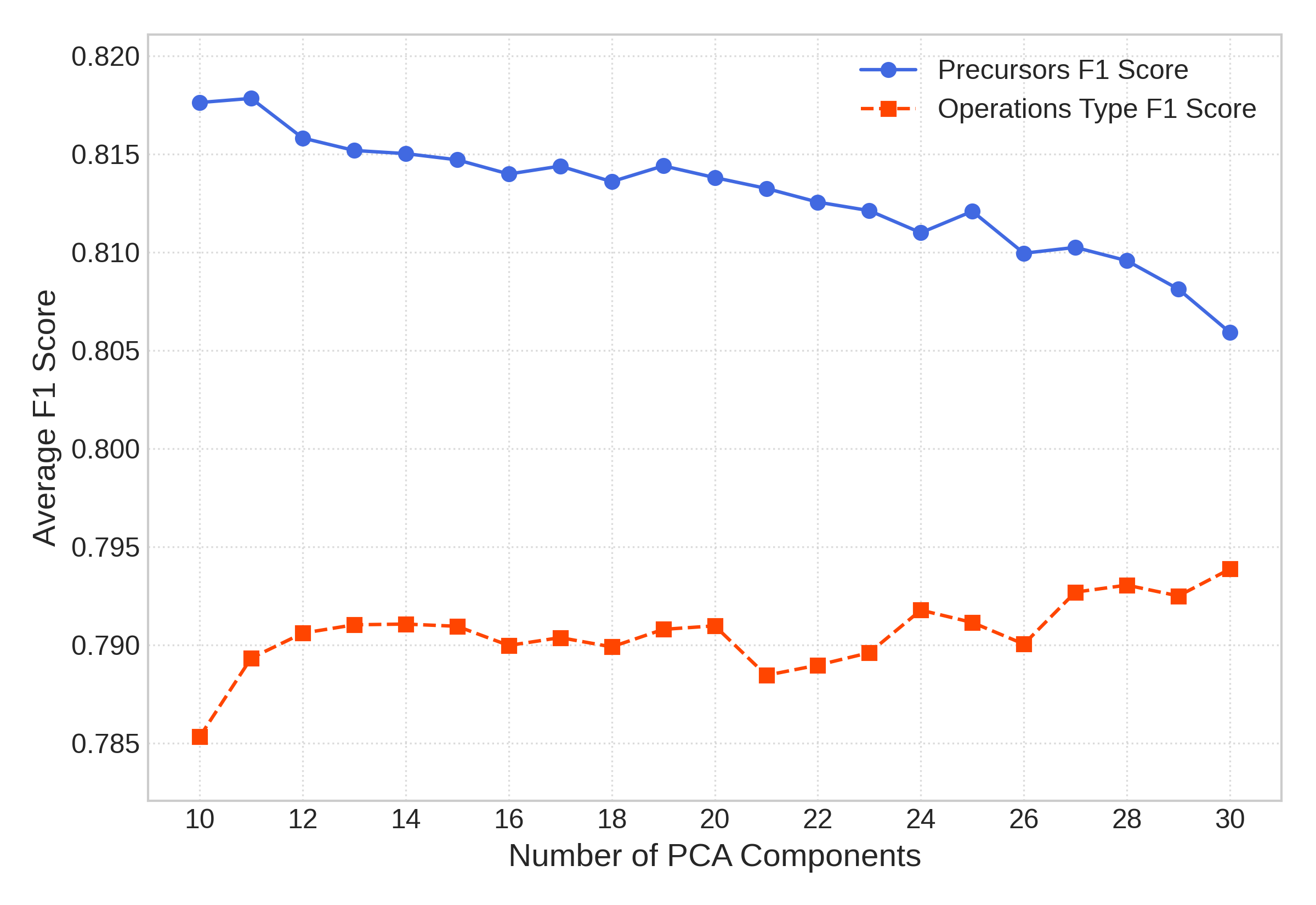}
    \caption*{(b) F1 scores vs. PCA components}
    \end{minipage}
    \caption{PCA analysis of \actiongraph adjacency matrices: (a) Individual and cumulative explained variance showing how information is distributed across components; (b) Trade-off between precursor and operation F1 scores as more structural information is incorporated.}
    \label{fig:pca_analysis}
\end{figure}

The PCA variance analysis (Figure~\ref{fig:pca_analysis}a) shows that structural information in \actiongraph adjacency matrices is distributed across many components, with the cumulative explained variance reaching approximately 50\% at 30 components. This indicates that synthesis graph structures are complex and no single component dominates the representation. The individual contribution of each component (blue bars) decreases gradually, suggesting that even later components contain meaningful information about synthesis pathways.

Figure~\ref{fig:pca_analysis}b reveals a clear trade-off as we increase the number of PCA components: precursor F1 score peaks at 10-11 components (0.818), after which it gradually declines, while operation F1 score improves with additional components, rising from 0.785 at 10 components to 0.794 at 30 components. This divergence suggests that precursor prediction benefits from a more composition-dominated representation, while operation prediction leverages richer structural encoding.

\begin{figure}[!htbp]
    \centering
    \begin{minipage}{0.46\textwidth}
    \centering
    \includegraphics[width=\linewidth]{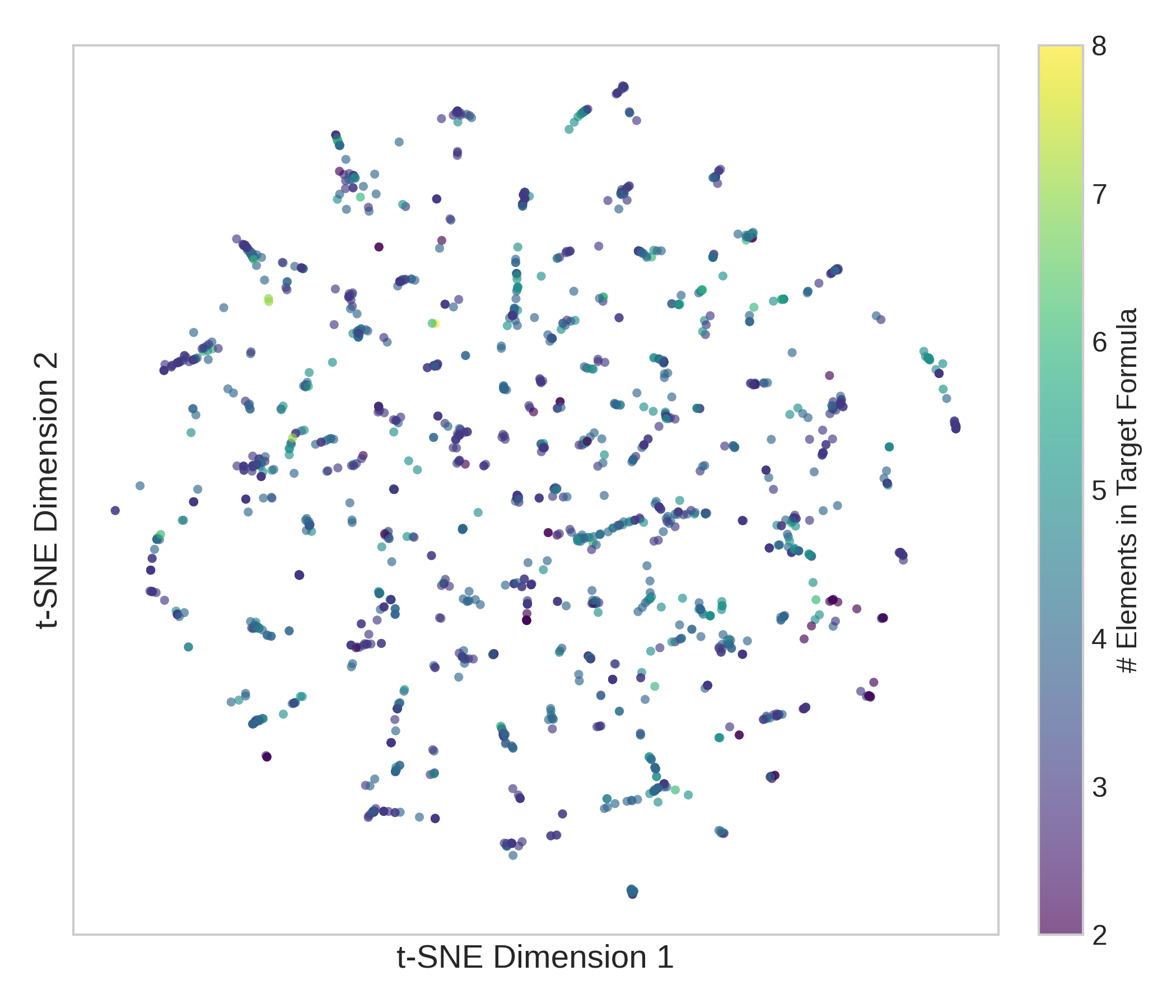}
    \caption*{(a) Baseline features}
    \end{minipage}
    \hfill
    \begin{minipage}{0.46\textwidth}
    \centering
    \includegraphics[width=\linewidth]{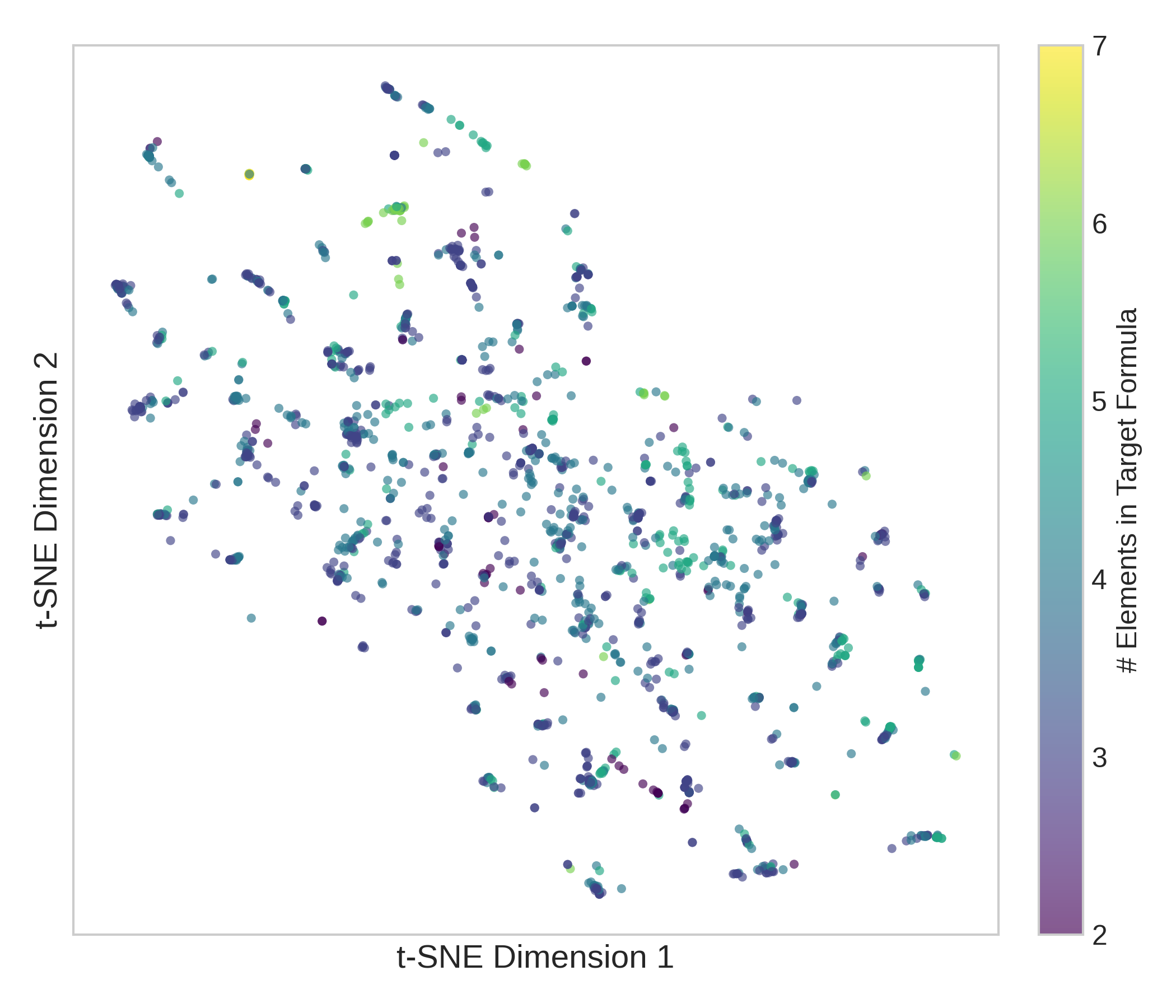}
    \caption*{(b) \actiongraph features (30 PCA components)}
    \end{minipage}
    \caption{t-SNE visualizations of feature spaces colored by number of elements in target formula: (a) Baseline features showing less defined clustering; (b) \actiongraph features showing more pronounced structure, indicating better encoding of synthesis similarities.}
    \label{fig:tsne_comparison}
\end{figure}

To qualitatively assess the effect of \actiongraph structural features, we visualized the feature spaces using t-SNE (Figure~\ref{fig:tsne_comparison}). The \actiongraph-based features (Figure~\ref{fig:tsne_comparison}b) produce more pronounced and structured clusters compared to the baseline representation (Figure~\ref{fig:tsne_comparison}a). This suggests that the inclusion of synthesis topology enables the model to group reactions with similar procedural flows, not just similar compositions.

\subsection{Quantitative Results and Discussion}

Table~\ref{tab:metrics} summarizes the key quantitative results for the baseline and \actiongraph-enhanced models at representative PCA component counts. The \actiongraph model with 10 PCA components (AG-PCA10) achieves the highest precursor F1 (0.818), while the 20-component model (AG-PCA20) yields the best operation F1 (0.791). Notably, operation length matching improves dramatically from 15.8\% (baseline) to 53.3\% (AG-PCA20), a 3.4$\times$ increase, indicating that structural information is critical for capturing the procedural fidelity of synthesis.

\begin{table}[htbp]
    \centering
    \caption{Performance comparison across models (test set $n=2,968$)}
    \label{tab:metrics}
    \begin{tabular}{lccc}
        \toprule
        Metric & Baseline & AG-PCA10 & AG-PCA20 \\
        \midrule
        \textbf{Precursors} & & & \\
        \hspace{2mm} F1 & 0.7995 & 0.8176 & 0.8138 \\
        \hspace{2mm} Exact Match & 0.509 & 0.567 & 0.550 \\
        \midrule
        \textbf{Operations} & & & \\
        \hspace{2mm} F1 & 0.763 & 0.785 & 0.791 \\
        \hspace{2mm} Length Match (\%) & 15.8 & 47.5 & 53.3 \\
        \bottomrule
        \end{tabular}
\end{table}

The observed divergence in F1 trends (Figure~\ref{fig:pca_analysis}b) highlights the importance of tuning the number of PCA components to balance precursor and operation prediction objectives. The optimal regime appears to be 10--15 components for precursor-centric tasks, and 20--30 for operation-centric tasks.

\textit{Note}: using different seeds for the train-test split did not significantly alter results.

\textbf{Structural analysis.} The PCA variance plot (Figure~\ref{fig:pca_analysis}a) confirms that no single component dominates the structural encoding; rather, meaningful information is distributed across many components, reflecting the complexity of inorganic synthesis pathways. The t-SNE visualizations further support the conclusion that \actiongraph features capture procedural similarities not accessible to composition-only models.

\textbf{Interpretability and limitations.} While \actiongraph encoding substantially improves procedural fidelity (operation sequence and length), exact operation matches remain below 9\%, reflecting the inherent ambiguity and sparsity of metadata in text-mined synthesis descriptions. Future work may address this gap by incorporating richer operation metadata or by learning hierarchical graph representations.

Overall, these results validate the core hypothesis: encoding synthesis structure via the \actiongraph framework enhances prediction of synthesis operations and improves procedural accuracy, representing a step toward robust inorganic synthesis pathway prediction from product formulas alone.

\textbf{Computational resources.} Experiments were conducted on an ASUS ROG Zephyrus G14 laptop with AMD Ryzen 9 4900HS processor and NVIDIA GeForce RTX 2060 Max-Q GPU. Each experiment completed in under one minute, with runtime increasing proportionally with the number of PCA components used.

\section{Conclusion and Outlook}\label{sec:conclusion}
In this work, we introduced the \actiongraph framework and demonstrated its effectiveness for predicting inorganic synthesis pathways given only the target product. Our approach encodes synthesis reactions as directed acyclic graphs and utilizes $k$-nearest neighbors with PCA-reduced adjacency matrices to capture structural information. The results show that incorporating \actiongraph structure improves both precursor and operation prediction compared to composition-only baselines.

A key insight from our experiments is the trade-off between precursor and operation prediction performance as a function of structural encoding depth. Precursor F1 scores peak at 10-11 PCA components before declining, while operation F1 scores continue improving with additional components. This suggests that precursor selection relies primarily on composition information, while operation prediction benefits from richer structural encoding. The dramatic improvement in operation length matching accuracy (from 15.8\% to 53.3\%) demonstrates that \actiongraph structural information is particularly valuable for capturing procedural aspects of synthesis.

Looking forward, several promising directions emerge. First, more sophisticated featurization methods could enhance both the chemical and structural representations in our model. While our current element fraction and property-based features provide a reasonable baseline, incorporating crystal structure information or learned embeddings could improve prediction accuracy~\citep{GSTVBL25}. Second, the \actiongraph framework could be extended to include more detailed operation conditions and intermediate phases, provided sufficient data becomes available. Third, alternative dimensionality reduction techniques or graph embedding methods might preserve more structural information than our current PCA approach.

The significant improvement in operation length matching suggests that the \actiongraph framework effectively capture synthesis procedure complexity, but the relatively modest gains in F1 scores indicate room for further refinement. Future work could explore dynamic weighting between compositional and structural features to optimize for specific prediction tasks.

We envision this framework becoming part of a closed-loop system where synthesis predictions guide automated experimentation, generating new data that further refines the model. Such a system could dramatically accelerate inorganic materials discovery by reducing reliance on trial-and-error approaches. While substantial challenges remain in data quality and model sophistication, the \actiongraph approach represents an important step toward data-driven synthesis planning for novel inorganic materials.

\clearpage

\bibliographystyle{alpha}
\bibliography{main}

@Article{elsamman2024,
author ="El-Samman, Amer Marwan and Husain, Incé Amina and Huynh, Mai and De Castro, Stefano and Morton, Brooke and De Baerdemacker, Stijn",
title  ="Global geometry of chemical graph neural network representations in terms of chemical moieties",
journal  ="Digital Discovery",
year  ="2024",
volume  ="3",
issue  ="3",
pages  ="544-557",
publisher = "Royal Society of Chemistry",
doi  ="10.1039/D3DD00200D",
url  ="http://dx.doi.org/10.1039/D3DD00200D"}

@article{kononova2019,
  author    = {Olga Kononova and Haoyan Huo and Tanjin He and Ziqin Rong and Tiago Botari and Wenhao Sun and Vahe Tshitoyan and Gerbrand Ceder},
  title     = {Text-mined dataset of inorganic materials synthesis recipes},
  journal   = {Scientific Data},
  year      = {2019},
  volume    = {6},
  number    = {1},
  pages     = {203},
  publisher = {Springer Science and Business Media LLC},
  month     = {Oct},
  doi       = {10.1038/s41597-019-0224-1}
}

@article{huo2022,
  author    = {Haoyan Huo and Christopher J. Bartel and Tanjin He and Amalie Trewartha and Alexander Dunn and Bin Ouyang and Anubhav Jain and Gerbrand Ceder},
  title     = {Machine-Learning Rationalization and Prediction of Solid-State Synthesis Conditions},
  journal   = {Chemistry of Materials},
  year      = {2022},
  volume    = {34},
  number    = {16},
  pages     = {7323--7336},
  publisher = {American Chemical Society},
  month     = {Aug},
  doi       = {10.1021/acs.chemmater.2c01293}
}

@misc{alabi2024empiredbdataacceleratecomputational,
      title={EmpireDB: Data System to Accelerate Computational Sciences}, 
      author={Daniel Alabi and Eugene Wu},
      year={2024},
      eprint={2412.10546},
      archivePrefix={arXiv},
      primaryClass={cs.DB},
      url={https://arxiv.org/abs/2412.10546}, 
}

@inproceedings{AGMSW25,
  author       = {Daniel Alabi and
                  Sainyam Galhotra and
                  Shagufta Mehnaz and
                  Zeyu Song and
                  Eugene Wu},
  title        = {Privacy and Security in Distributed Data Markets},
  booktitle    = {Companion of the 2025 International Conference on Management of Data,
                  {SIGMOD/PODS} 2025, Berlin, Germany, June 22-27, 2025},
  publisher    = {{ACM}},
  year         = {2025}
}

@article{szymanski_toward_2021,
	title = {Toward autonomous design and synthesis of novel inorganic materials},
	volume = {8},
	issn = {2051-6347, 2051-6355},
	url = {https://xlink.rsc.org/?DOI=D1MH00495F},
	doi = {10.1039/D1MH00495F},
	language = {en},
	number = {8},
	urldate = {2025-05-06},
	journal = {Materials Horizons},
	author = {Szymanski, Nathan J. and Zeng, Yan and Huo, Haoyan and Bartel, Christopher J. and Kim, Haegyeom and Ceder, Gerbrand},
	year = {2021},
	pages = {2169--2198}
}

@article{kim_predicting_2024,
	title = {Predicting synthesis recipes of inorganic crystal materials using elementwise template formulation},
	volume = {15},
	issn = {2041-6520, 2041-6539},
	url = {https://xlink.rsc.org/?DOI=D3SC03538G},
	doi = {10.1039/D3SC03538G},
	language = {en},
	number = {3},
	urldate = {2025-04-02},
	journal = {Chemical Science},
	author = {Kim, Seongmin and Noh, Juhwan and Gu, Geun Ho and Chen, Shuan and Jung, Yousung},
	year = {2024},
	pages = {1039--1045}
}

@article{carleo2017solving,
  title={Solving the quantum many-body problem with artificial neural networks},
  author={Carleo, Giuseppe and Troyer, Matthias},
  journal={Science},
  volume={355},
  number={6325},
  pages={602--606},
  year={2017}
}

@article{mills2021role,
  title={The role of machine learning in scientific simulations},
  author={Mills, Kyle and Spanner, Michael and Tamblyn, Isaac},
  journal={Nature Reviews Physics},
  volume={3},
  number={6},
  pages={447--460},
  year={2021}
}

@article{nieuwenburg2017learning,
  title={Learning phase transitions by confusion},
  author={van Nieuwenburg, Evert PL and Liu, Yi-Hong and Huber, Sebastian D},
  journal={Nature Physics},
  volume={13},
  number={5},
  pages={435--439},
  year={2017}
}

@article{raissi2019physics,
  title={Physics-informed neural networks: A deep learning framework for solving forward and inverse problems involving nonlinear partial differential equations},
  author={Raissi, Maziar and Perdikaris, Paris and Karniadakis, George E},
  journal={Journal of Computational Physics},
  volume={378},
  pages={686--707},
  year={2019}
}

@inproceedings{gilmer2017neural,
  title={Neural message passing for quantum chemistry},
  author={Gilmer, Justin and Schoenholz, Samuel S and Riley, Patrick F and Vinyals, Oriol and Dahl, George E},
  booktitle={Proceedings of the 34th International Conference on Machine Learning (ICML)},
  year={2017}
}

@article{zhavoronkov2019deep,
  title={Deep learning enables rapid identification of potent DDR1 kinase inhibitors},
  author={Zhavoronkov, Alex and Ivanenkov, Yan A and Aliper, Alex and Veselov, Maksim and Aladinskiy, Vladimir and Aladinskaya, Anastasiya and Terentiev, Victor A and Polykovskiy, Daniil and Kuznetsov, Mikhail and Asadulaev, Arman and others},
  journal={Nature Biotechnology},
  volume={37},
  number={9},
  pages={1038--1040},
  year={2019}
}

@article{gomez2018automatic,
  title={Automatic chemical design using a data-driven continuous representation of molecules},
  author={G{\'o}mez-Bombarelli, Rafael and Wei, Jennifer N and Duvenaud, David and Hern{\'a}ndez-Lobato, Jos{\'e} Miguel and S{\'a}nchez-Lengeling, Benjamin and Sheberla, Dennis and Aguilera-Iparraguirre, Jorge and Hirzel, Timothy D and Adams, Ryan P and Aspuru-Guzik, Al{\'a}n},
  journal={ACS Central Science},
  volume={4},
  number={2},
  pages={268--276},
  year={2018}
}

@article{schutt2018schnet,
  title={SchNet: A continuous-filter convolutional neural network for modeling quantum interactions},
  author={Sch{\"u}tt, Kristof T and Sauceda, Huziel E and Kindermans, Pieter-Jan and Tkatchenko, Alexandre and M{\"u}ller, Klaus-Robert},
  journal={NeurIPS},
  year={2018}
}

@article{battaglia2018relational,
  title={Relational inductive biases, deep learning, and graph networks},
  author={Battaglia, Peter and Hamrick, Jessica and Bapst, Victor and Sanchez-Gonzalez, Alvaro and Zambaldi, Vinicius and Malinowski, Mateusz and Tacchetti, Andrea and Raposo, David and Santoro, Adam and Faulkner, Ryan and others},
  journal={arXiv preprint arXiv:1806.01261},
  year={2018}
}

@article{jumper2021highly,
  title={Highly accurate protein structure prediction with AlphaFold},
  author={Jumper, John and Evans, Richard and Pritzel, Alexander and Green, Tim and Figurnov, Michael and Ronneberger, Olaf and Tunyasuvunakool, Kathryn and Bates, Russ and {\v{Z}}{\'\i}dek, Augustin and Potapenko, Anna and others},
  journal={Nature},
  volume={596},
  number={7873},
  pages={583--589},
  year={2021}
}

@article{sanchez2018inverse,
  title={Inverse molecular design using machine learning: Generative models for matter engineering},
  author={Sanchez-Lengeling, Benjamin and Aspuru-Guzik, Al{\'a}n},
  journal={Science},
  volume={361},
  number={6400},
  pages={360--365},
  year={2018}
}

@article{popova2018deep,
  title={Deep reinforcement learning for de novo drug design},
  author={Popova, Mariya and Isayev, Olexandr and Tropsha, Alexander},
  journal={Science Advances},
  volume={4},
  number={7},
  pages={eaap7885},
  year={2018}
}

@article{rolnick2019tackling,
  title={Tackling climate change with machine learning},
  author={Rolnick, David and Donti, Priya L and Kaack, Lynn H and Kochanski, Kelly and Lacoste, Alexandre and Sankaran, Krishna and Ross, Andrew S and Milojevic-Dupont, Nicolas and Jaques, Natasha and Waldman-Brown, Anna and others},
  journal={arXiv preprint arXiv:1906.05433},
  year={2019}
}

@article{liu2017accelerated,
  title={Accelerated Monte Carlo simulations using restricted Boltzmann machines},
  author={Liu, Jun and Wang, Lei},
  journal={Physical Review B},
  volume={96},
  number={14},
  pages={144426},
  year={2017}
}

@article{macleod2020self,
  title={Self-driving laboratory for accelerated discovery of thin-film materials},
  author={MacLeod, Benjamin P and Parlane, Fraser G L and Morrissey, Thomas D and Häse, Florian and Roch, Loïc M and Dettelbach, Kevan E and Moreira, Rafael and Yunker, Peter J and Aspuru-Guzik, Alán and Hein, Jason E and others},
  journal={Science Advances},
  volume={6},
  number={20},
  pages={eaaz8867},
  year={2020}
}

@article{he_precursor_2023,
	title = {Precursor recommendation for inorganic synthesis by machine learning materials similarity from scientific literature},
	volume = {9},
	issn = {2375-2548},
	url = {http://arxiv.org/abs/2302.02303},
	doi = {10.1126/sciadv.adg8180},
	number = {23},
	urldate = {2025-05-11},
	journal = {Science Advances},
	author = {He, Tanjin and Huo, Haoyan and Bartel, Christopher J. and Wang, Zheren and Cruse, Kevin and Ceder, Gerbrand},
	month = jun,
	year = {2023},
	note = {arXiv:2302.02303 [cond-mat]},
	pages = {eadg8180}
}

@article{kim_large_2024,
	title = {Large {Language} {Models} for {Inorganic} {Synthesis} {Predictions}},
	volume = {146},
	copyright = {https://doi.org/10.15223/policy-029},
	issn = {0002-7863, 1520-5126},
	url = {https://pubs.acs.org/doi/10.1021/jacs.4c05840},
	doi = {10.1021/jacs.4c05840},
	language = {en},
	number = {29},
	urldate = {2025-05-11},
	journal = {Journal of the American Chemical Society},
	author = {Kim, Seongmin and Jung, Yousung and Schrier, Joshua},
	month = jul,
	year = {2024},
	pages = {19654--19659}
}

@misc{karpovich_inorganic_2021,
	title = {Inorganic {Synthesis} {Reaction} {Condition} {Prediction} with {Generative} {Machine} {Learning}},
	url = {http://arxiv.org/abs/2112.09612},
	doi = {10.48550/arXiv.2112.09612},
    urldate = {2025-05-11},
	publisher = {arXiv},
	author = {Karpovich, Christopher and Jensen, Zach and Venugopal, Vineeth and Olivetti, Elsa},
	month = dec,
	year = {2021},
	note = {arXiv:2112.09612 [cond-mat]}
}

@article{mcdermott_graph-based_2021,
	title = {A graph-based network for predicting chemical reaction pathways in solid-state materials synthesis},
	volume = {12},
	issn = {2041-1723},
	url = {https://www.nature.com/articles/s41467-021-23339-x},
	doi = {10.1038/s41467-021-23339-x},
	language = {en},
	number = {1},
	urldate = {2025-04-01},
	journal = {Nature Communications},
	author = {McDermott, Matthew J. and Dwaraknath, Shyam S. and Persson, Kristin A.},
	month = may,
	year = {2021},
	pages = {3097},
}

@article{wei_machine_2025,
	title = {Machine learning-assisted retrosynthesis planning: {Current} status and future prospects},
	volume = {77},
	issn = {10049541},
	shorttitle = {Machine learning-assisted retrosynthesis planning},
	url = {https://linkinghub.elsevier.com/retrieve/pii/S1004954124003720},
	doi = {10.1016/j.cjche.2024.10.014},
	language = {en},
	urldate = {2025-05-15},
	journal = {Chinese Journal of Chemical Engineering},
	author = {Wei, Yixin and Shan, Leyu and Qiu, Tong and Lu, Diannan and Liu, Zheng},
	month = jan,
	year = {2025},
	pages = {273--292},
}

@article{GSTVBL25,
	author = {Guo, Gabe and Saidi, Tristan Luca and Terban, Maxwell W. and Valsecchi, Michele and Billinge, Simon J. L. and Lipson, Hod},
	date = {2025/04/28},
	doi = {10.1038/s41563-025-02220-y},
	id = {Guo2025},
	isbn = {1476-4660},
	journal = {Nature Materials},
	title = {Ab initio structure solutions from nanocrystalline powder diffraction data via diffusion models},
	url = {https://doi.org/10.1038/s41563-025-02220-y},
	year = {2025}}

@article{BP24,
author = {Billinge, Simon J. L. and Proffen, Thomas},
title = {Machine learning in crystallography and structural science},
journal = {Acta Crystallographica Section A},
volume = {80},
number = {2},
pages = {139-145},
doi = {https://doi.org/10.1107/S2053273324000172},
url = {https://onlinelibrary.wiley.com/doi/abs/10.1107/S2053273324000172},
eprint = {https://onlinelibrary.wiley.com/doi/pdf/10.1107/S2053273324000172},
year = {2024}
}

@article{BL07,
author = {Simon J. L. Billinge  and Igor Levin },
title = {The Problem with Determining Atomic Structure at the Nanoscale},
journal = {Science},
volume = {316},
number = {5824},
pages = {561-565},
year = {2007},
doi = {10.1126/science.1135080},
URL = {https://www.science.org/doi/abs/10.1126/science.1135080},
eprint = {https://www.science.org/doi/pdf/10.1126/science.1135080}}

@ARTICLE{2008AcCrA..64..631J,
       author = {{Juhas}, P. and {Granlund}, L. and {Duxbury}, P.~M. and {Punch}, W.~F. and {Billinge}, S.~J.~L.},
        title = "{The Liga algorithm for ab initio determination of nanostructure}",
      journal = {Acta Crystallographica Section A},
         year = 2008,
        month = nov,
       volume = {64},
        pages = {631-640},
          doi = {10.1107/S0108767308027591}
}

@article{NZTB25,
	author = {Na Narong, Tanaporn and Zachko, Zoe N. and Torrisi, Steven B. and Billinge, Simon J. L.},
	doi = {10.1038/s41524-025-01589-3},
	isbn = {2057-3960},
	journal = {npj Computational Materials},
	number = {1},
	pages = {98},
	title = {Interpretable multimodal machine learning analysis of X-ray absorption near-edge spectra and pair distribution functions},
	url = {https://doi.org/10.1038/s41524-025-01589-3},
	volume = {11},
	year = {2025}}

@article{scikit-learn,
  title={Scikit-learn: Machine Learning in {P}ython},
  author={Pedregosa, F. and Varoquaux, G. and Gramfort, A. and Michel, V.
          and Thirion, B. and Grisel, O. and Blondel, M. and Prettenhofer, P.
          and Weiss, R. and Dubourg, V. and Vanderplas, J. and Passos, A. and
          Cournapeau, D. and Brucher, M. and Perrot, M. and Duchesnay, E.},
  journal={Journal of Machine Learning Research},
  volume={12},
  pages={2825--2830},
  year={2011}
}

@ARTICLE{Cover13,
  author={Cover, T. and Hart, P.},
  journal={IEEE Transactions on Information Theory}, 
  title={Nearest neighbor pattern classification}, 
  year={1967},
  volume={13},
  number={1},
  pages={21-27},
  doi={10.1109/TIT.1967.1053964}}

@misc{noh2025,
      title={Retrieval-Retro: Retrieval-based Inorganic Retrosynthesis with Expert Knowledge}, 
      author={Heewoong Noh and Namkyeong Lee and Gyoung S. Na and Chanyoung Park},
      year={2025},
      eprint={2410.21341},
      archivePrefix={arXiv},
      primaryClass={cs.LG},
      url={https://arxiv.org/abs/2410.21341}, 
}

\end{document}